# Effect of gas pressure on plasma asymmetry and higher harmonics generation in sawtooth waveform driven capacitively coupled plasma discharge

Sarveshwar Sharma[1,2], Miles Turner[3], and Nishant Sirse[4]

[1]Institute for Plasma Research, Bhat, Gandhinagar, Gujarat, 382428 India
[2]Homi Bhabha National Institute, Training School Complex, Anushaktinagar, Mumbai-400094, India
[3]School of Physical Sciences and National Centre for Plasma Science and Technology, Dublin City University, Dublin-D9, Ireland
[4]Institute of Science and Research and Centre for Scientific and Applied Research, IPS Academy, Indore-452012, India

E-mail: nishantsirse@ipsacademy.org, sarvesh@ipr.res.in

**Abstract**

Using particle-in-cell (PIC) simulation technique, the effect of gas pressure (5 – 500 mTorr) on the plasma spatial asymmetry, ionization rate, metastable gas densities profile, electron energy distribution function and higher harmonics generation are studied in a symmetric capacitively coupled plasma discharge driven by a sawtooth-like waveform. At a constant current density of 50 A/m$^2$, the simulation results predict a decrease in the plasma spatial asymmetry (highest at 5mTorr) with increasing gas pressure reaching a minimum value (at intermediate gas pressures) and then turning into a symmetric discharge at higher gas pressures. Conversely, the flux asymmetry shows an opposite trend. At a low gas pressure, the observed strong plasma spatial asymmetry is due to high frequency oscillation on the instantaneous sheath edge position near to one of the electrodes triggered by temporally asymmetry waveform, whereas the flux asymmetry is not present due to collision-less transport of charge particles. At higher pressures, multi-step ionization through metastable states dominates in the plasma bulk, causing a reduction in the plasma spatial asymmetry. Distinct higher harmonics (26$^{th}$) are observed in the bulk electric field at low pressure and diminished at higher gas pressures. The electron energy distribution function changes its shape from bi-Maxwellian at 5 mTorr to nearly Maxwellian at intermediate pressures and then depletion of the high-energy electrons (below 25 eV) is observed at higher gas pressures. The inclusion of the secondary electron emission is found to be negligible on the observed simulation trend.

1. **Introduction**

Capacitively coupled plasmas (CCPs) are among the most essential etching tools in the semiconductor industry for fabricating large-scale integrated circuits[1]. Higher ion flow improves processing rate based on plasma density, whereas lesser energy prevents damage to the substrate. To generate higher densities in CCP discharges with single-frequency (SF) sinusoidal waveforms, use larger voltages or higher frequencies while keeping other operating parameters unchanged.[1-11]

Currently, there is significant interest in operating CCPs at frequencies as high as a few tens of MHz. Compared to low-frequency plasma operation (such as the traditional 13.56 MHz), a CCP operating in the very high frequency (VHF) band (30-300 MHz) generates higher plasma density due to increased discharge current for the selected discharge power and reduced DC self-bias, resulting in high-rate, low-damage processing.[12-14] In addition to plasma density and temperature, the plasma process is influenced by radical densities, which are estimated by the *Electron Energy Distribution Function* (EEDF). Specifically, the electron population above the threshold energy leads to electron-impact dissociation of gas molecules. Consequently, managing the EEDF is critical for improving and optimizing plasma processing rates. Earlier experimental research has revealed a significant shift in the EEDF with changes in the driving frequency. Abdel-Fattah *et al.*[15-16] found that at a 13.56 MHz operating frequency, a bi-Maxwellian (or convex-type for non-Ramseur gases) EEDF is observed, while VHF plasma excitation typically exhibits a bi-Maxwellian EEDF regardless of gas pressure or type (Ramsauer or non-Ramsauer). This is due to the increased stochastic heating in VHF-triggered CCP. Simulation studies have also shown the impact of driving frequency (27.12–70 MHz) on the EEDF and electron-sheath interactions in low-pressure capacitive discharges using self-consistent particle-in-cell/Monte Carlo collision (PIC/MCC) methods.[5,7,9,17,18] Scientific research on very-high-frequency CCP discharges also describes features such as electric field transients and their impact on plasma and sheaths, as well as the presence of higher harmonics in voltage and current profiles[6,7,19-21].

Single-frequency CCP driven by VHF in the presence of a weak, uniform transverse magnetic field is another novel technique, where a significant increase in plasma density has been observed.[22-24] Patil et al.[22] were the first to demonstrate that the enhanced performance is due to a resonance effect, occurring when the electron cyclotron frequency ($f_{ce} = eB/m_e$, where e, $m_e$, and B are the electronic charge, mass, and external applied magnetic field, respectively) equals half of the applied RF frequency ($f_{rf}$). Zhang *et.al*[23] observed similar results using PIC simulations in the frequency range of 13.56-60 MHz and confirmed their findings experimentally. Subsequently, Sharma et al.[24] demonstrated that the tail-end electron population is highest in the resonance case using the EEDF profile. Asymmetry in flux, density, and sheath is also observed in low-pressure, weakly magnetized SF-CCP discharges.[25]

Dual-frequency CCP discharge is also an alternate technique to create high-density plasmas either by varying the high-frequency voltage amplitude, or the higher frequency itself gives an additional grip on the ion energy by adjusting the lower frequency parameters[26]. CCP discharges driven by DF waveforms under various operating conditions (i.e. by varying voltage, pressure, frequency etc.) have been extensively studied in the literature through both simulations and experiments.[27-32] One of the research outcomes revealed that applying an intermediate frequency in DF-CCPs can influence the ion energy distribution function (IEDF) [33]. In this particular kind of triple frequency driven CCP arrangement, the influence of intermediate frequency on plasma properties has been explored using various numerical simulation approaches.[33-37]

In recent studies, tailored waveform driven CCP discharges have shown promising results in terms of plasma asymmetry generation, and to control the ion energy and ion flux on the electrodes.[38-50] In such types of waveforms, higher harmonics are superimposed on sinusoidal waveform for modifying either its amplitude or slope to produce a non-sinusoidal shape. These non-sinusoidal waveforms generate an asymmetric sheath response on the powered and grounded electrodes, which is responsible for an uneven power deposition to the plasma electrons through RF sheaths. Phase and amplitude asymmetry has been studied thoroughly for multiple sinusoidal frequency excitation,[39-42] and higher harmonics superimposed non-sinusoidal waveform such as Gaussian and peaks/valleys waveform.[38,43-44] The slope asymmetry in CCP discharges was introduced by using a sawtooth-like waveform that possesses different rise and fall time[45]. In an experimental study using phase resolved optical emission spectroscopy (PROES) performed by Bruneau et al.[45-47] for voltage driven CCP discharge showed the sawtooth-like waveform produces strong ionization asymmetry. The electrode with a faster sheath expansion creates more ionization whereas the electrode with slower sheath expansion generates less ionization. Later using PIC simulation,[48] the effect of driving frequency on discharge asymmetry, electric field nonlinearity and electron heating mechanism in a low pressure CCP discharge was investigated. The simulation results predict higher frequencies oscillations on the instantaneous sheath edge position at lower driving frequencies (13.56 MHz), whereas the high frequency (27.12 MHz and 54.24 MHz) causes electric field oscillation in the bulk plasma due to electron beams generation from near to the sheath region. However, spatial asymmetry is observed to decrease with increasing driving frequency. On the other hand, the ion energy distribution function (IEDF) study[48] shows narrow low-energy peak at higher driving frequencies. Sharma et al.[49] performed a systematic simulation study by varying the number of harmonics contained in the saw-tooth like waveform to investigate the plasma asymmetry. The results showed that by varying the number of harmonics the ion flux asymmetry is not altered, whereas the sheath asymmetry and thus the ion energy could be significantly modified. The IEDF showed a transition from a broad bi-modal shape to a narrow-shaped distribution. In earlier studies, the phenomenon of field reversal and charge separation in single and dual-frequency CCP discharges has been reported[29,32,50-55]. More recently,[56] field reversal phenomenon, charge separation, and electron and ion heating were investigated in a CCP discharge driven by a sawtooth-like waveform. The results predict the formation of localized multiple field reversal regions during both expanding and collapsing phases of the sheath edge. These multiple field reversal regions are formed by the charge separation that is also responsible for the efficient electron heating and ion cooling near the sheath region. In the present article, we follow our previous studies[49,56-58] and investigate the effect of gas pressure on the spatial asymmetry and high harmonics generation in a CCP discharge excited by a sawtooth-like current waveform. The present simulation studies are performed for a saw-tooth current waveform withan ideal sawtooth-like waveform that contains large number of harmonics ($N = 50$).

The article is structured as follows. The simulation technique, assumptions and parametric conditions are presented in section II. The physical interpretation and description of simulation results are discussed in section III. In section IV, the summary and conclusion of the work are given.

## 2. Simulation Technique and Parameters

Here, a very well-benchmarked and tested 1D3V electrostatic self-consistent particle-in-cell (PIC) code is used to model a current-driven symmetric capacitively coupled discharge in argon plasma. The foundation of this simulation methodology is the Particle-in-Cell/Monte Carlo collision (PIC/MCC) methods, which are detailed in literature.[59-60] Prof. Miles Turner developed this code at Dublin City University, and it is widely utilized in a number of academic articles, few of which are available here.[6, 8, 17, 37, 61-64] For this study, we have chosen to execute this code in the *current driven* manner, while it can also operate in *voltage driven* mode. The simulation technique's specifics are documented in published works[18, 65-66]. All significant particle collision reactions, like ion-neutral (inelastic, elastic, and charge exchange) and electron-neutral (inelastic, elastic, and ionisation) are considered here. Our simulation has also considered other significant reactions, such as metastable pooling, multi-step ionisation, super elastic collisions, partial de-excitation, and further de-excitation. The research literature provides information on reaction types as well as additional details[8]. In the computer simulation, charged particles such as electrons and ions are included with two lumped excited states of Ar, namely Ar* ($3p^54s$), 11.6 eV, and Ar** ($3p^54p$), 13.1 eV, in a uniform neutral argon gas environment. In our output diagnostics, we have also tracked the formation of metastables (Ar*, Ar**) that were taken into account in the simulation.

The species information, reactions, and collision cross sections for the simulation were obtained from reliable sources in the scientific literature.[8, 62, 67] Selecting an appropriate grid size ($\Delta x$) and time step size ($\Delta t$) to resolve the electron plasma frequency $\left(f_{pe} = \left(\sqrt{n_0 e^2/\varepsilon_0 m_e}\right)/2\pi\right)$ and Debye length $\left(\lambda_{De} = \sqrt{\varepsilon_0 T_e/n_0 e}\right)$, respectively, ensures the stability and accuracy of the PIC method. The two electrodes are assumed to be flat, parallel to each other, and of infinite dimensions. Additionally, it is considered that the electrodes are fully absorbing, with secondary electron emission (both electron and ion) being neglected for the sake of simplicity. The gap between electrodes is 60 mm, and the system operates at an argon gas pressure ranging from 5 to 500 mTorr. The ions are at the same temperature as the neutral gas, which is evenly distributed at a constant temperature of 300 K. Each simulation scenario has 100 particles per cell. The simulations were conducted for over 6000 RF cycles to reach a steady state.

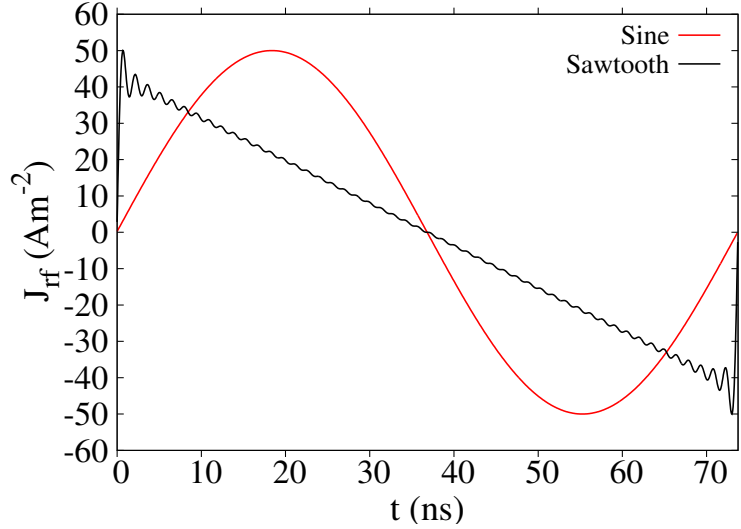

**Figure 1:** Waveform profiles, including sinusoidal and sawtooth, at 13.56 MHz with an amplitude of 50 A/m²

Complex waveform shapes can be generated by applying multiple harmonics of a fundamental RF frequency and carefully adjusting the phase shifts between them. To create the plasma, we adopted a sawtooth-like current waveform, as theoretically expressed by the following equation (see figure 1):

$$J_{rf}(t) = \pm J_0 \sum_{k=1}^{N} \frac{1}{k} \sin(k\omega_{rf} t) \quad \text{---------(1)}$$

where positive and negative signs indicate the "sawtooth-down" and "sawtooth-up" waveforms, respectively. Here, $J_o$ represents the amplitude of the current density at the powered electrode, while $\omega_{rf}$ denotes the fundamental angular frequency of the operating radiofrequency. The magnitude $J_o$ depends on the total number of harmonics, denoted as N (50 in this case), and is adjusted to achieve the desired peak-to-peak current density amplitude. For all our simulations, we utilized a "sawtooth-down" current waveform. The total current density amplitude ($J_o$) was kept constant at 50 A/m² in all simulations, and the fundamental frequency was consistently maintained at 13.56 MHz.

Figure 1 demonstrates that the current profile displays temporal symmetry for the sinusoidal waveform i.e. N =1. However, for N=50, the rise time (from 0 to 50 A/m²) is significantly reduced to nearly 0.72 ns. The current waveform is applied to the powered electrode at $L$=0 mm, and the grounded electrode is at $L$=60 mm. The simulations do not take into account any external capacitors. The DC self-bias at the powered electrode results from the temporal asymmetry[46-49,56-57] which arises self-consistently in the simulation due to the charge imbalance between the powered and grounded electrodes. It is further discussed in the next section.

## 3. Results and Discussions

We first investigate the effect of gas pressures on the plasma density profile including metastables ($Ar^*$ and $Ar^{**}$) gas densities in the discharge. Figure 2 (a), (b) and (c) shows the plot of plasma (electron and ion), $Ar^*$ and $Ar^{**}$ density respectively versus position where gas pressure varied from 5 mTorr to 500 mTorr. At 5 mTorr gas pressure, the ion and electron density profile plotted in figure 2 (a) shows a maximum in the centre of the discharge and decreases toward the electrodes. The peak plasma density in this case is ~$5\times10^{15}$ $m^{-3}$ observed at the centre of the discharge. As shown in figure 2 (b) and 2 (c), the density profile of $Ar^*$ and $Ar^{**}$ is similar to that of plasma density profile i.e., it is maximum ($Ar^*$ ~$8.6\times10^{15}$ $m^{-3}$ and $Ar^{**}$ ~$1.6\times10^{16}$ $m^{-3}$) at the centre of the discharge and falls near to the electrodes. However, the sheath asymmetry is highest at 5 mTorr gas pressure (figure 2 (d) and 2 (e)). The estimated sheath width at 5 mTorr is ~5.6 mm near to the grounded electrode, whereas it is ~4.2 mm near to the powered electrode i.e., a 33% asymmetry is observed. Here, the sheath dimensions are calculated from the simulation data by plotting the instantaneous electron density profiles within RF phase and finding the maximum distance from the electrode where quasi-neutrality condition breaks down. At 5 mTorr, the temporal asymmetric sawtooth-like waveform generates an approximate 46 V potential difference between plasma potential (~84 V) and DC self-bias (~38 V) obtained from the potential profile. Due to low pressure condition (5 mTorr), this will generate a distinguished single ion energy peak, which is observed in the simulation on both powered (46 eV) and grounded (84 eV) electrodes corresponding to their potential difference. On the other hand, the flux impinging on the powered and grounded electrode is nearly the same i.e., ~$5\times10^{18}$ $m^{-2}s^{-1}$. Figure 2 (d) and 2 (e) shows corresponding time-averaged electron/ion flux distribution in the discharge and sheath/flux asymmetry respectively versus gas pressure. It is clear from the simulation results that the low-pressure discharge generates sheath and ion energy asymmetry, but the flux asymmetry is not present. At a low gas pressure (5 mTorr), the symmetric density profiles and nearly equal flux on both powered and grounded electrodes are caused by the non-local electron behavior of electron impact processes due to longer electron mean free path (~3 cm). In this case, the high energy electrons after gaining the energy from oscillating sheath edge return into bulk plasma generating charged particles and excited state densities through ionization and excitation mechanisms respectively. The sheath asymmetry and thus difference in the ion energy arriving at different electrodes, is attributed to the temporally asymmetric sawtooth-like waveform, which produces a different sheath response at powered and grounded electrodes. The powered electrode sheath is expanding and collapsing slowly, whereas the grounded electrode is varying rapidly thus creating asymmetric plasma heating and flux imbalance.

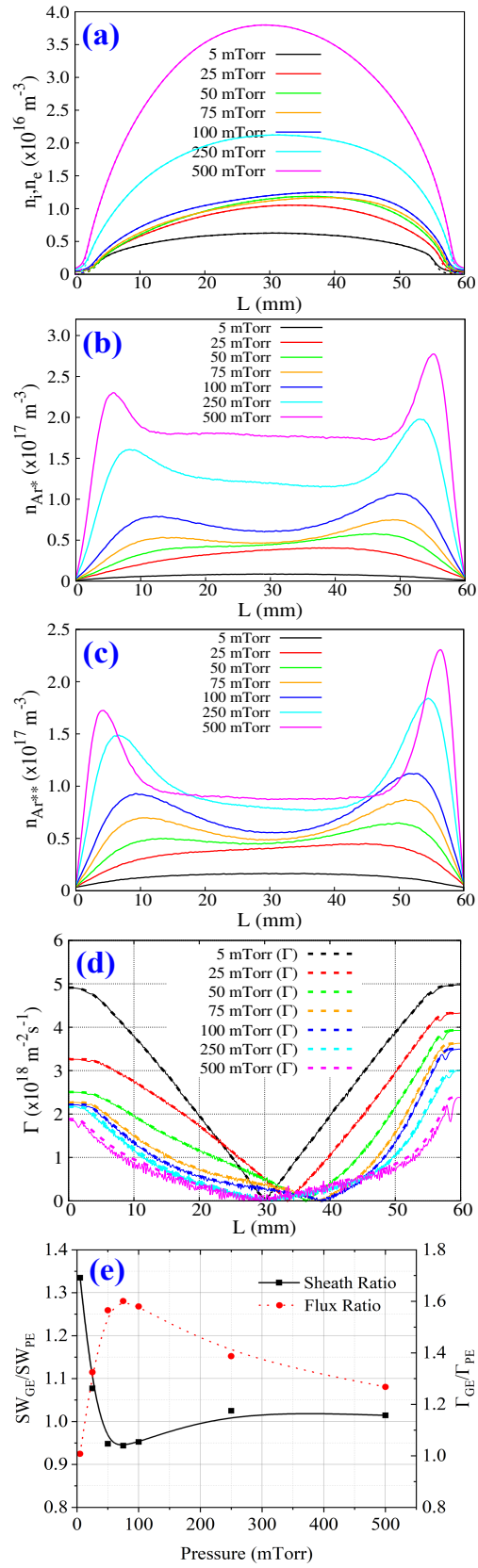

Figure 2: Time-averaged spatial profile of (a) electron and ion density, (b) Ar* density and (c) Ar** density (d) Time-averaged electron (line) and ion flux (dot) profile and (e) Sheath width

and flux ratio at the electrodes at different gas pressures from 5 mTorr to 500 mTorr. The current density and driving frequency are 50 A/m$^2$ and 27.12 MHz respectively.

As the gas pressure increases, it is observed that the maximum of the plasma density (figure 2 (a)) is shifting towards the grounded electrode generating spatial asymmetry. The Ar$^*$ (figure 2 (b)) and Ar$^{**}$ (figure 2 (c)) density profile showing 2 peaks, one near to the powered electrode and second near the grounded electrode. However, the metastables peak density near to the grounded electrode is higher when compared to the powered electrode. The shifting of plasma density near grounded and excitation localization near to sheath boundaries is attributed to reduction in electron mean free path due to increased gas pressure. Thus, the high energy electrons produced by the sheath interaction dissipate their energy near to the sheath edge producing local ionization and excitation. The effect is higher near to the grounded electrode because the applied temporally asymmetric sawtooth-like waveform is producing asymmetric sheath response at the electrodes. An increasing gas pressure is also affecting the sheath asymmetry. As shown in figure 2 (e), the sheath asymmetry is lowest around 75 mTorr gas pressure where the powered and grounded sheath widths are ~3.56 mm and ~3.36 mm respectively. On the other hand, the simulated ion flux at powered electrode ($\Gamma_{PE}$) is ~2.3×10$^{18}$ m$^{-2}$ s$^{-1}$, whereas flux at the grounded electrode ($\Gamma_{GE}$) is ~3.6×10$^{18}$ m$^{-2}$ s$^{-1}$ at 75 mTorr gas pressure. For the present operating parameters, these flux values correspond to the highest flux asymmetry of ~1.6 ($\Gamma_{GE}/\Gamma_{PE}$) i.e., 60% higher ion flux on the grounded electrode when compared to the powered electrode (figure 2 (e)). The DC self-bias and plasma potential continue to decrease with increasing gas pressure reaching to 13 V and 39 V respectively at 500 mTorr gas pressure. A variation in flux asymmetry with increasing gas pressure is associated with the asymmetric ionization rates, which is discussed later in the paper. Interestingly, it is observed that the density profile is turning symmetric at 250 and 500 mTorr gas pressure. This effect is attributed to an enhanced ionization in the bulk plasma from metastable states (multistep ionization) that plays a crucial role in the high-pressure regime. As shown in Figure 2 (b) and (c), the maximum density of Ar$^*$ and Ar$^{**}$ is reaching ~ (2-3) ×10$^{17}$ m$^{-3}$ at 500 mTorr gas pressure near the grounded electrode, which is approximately 10 times higher than the plasma density in the bulk plasma (Figure 2 (a)). Thus, the multistep ionization from Ar$^*$ and Ar$^{**}$ may play a significant role in the overall plasma density generation. This is discussed in the subsequent paragraph.

Figure 3 (a), (b) and (c) shows the time-averaged spatial profile of ionization rate from ground state, first excited state (Ar$^*$ - 11.6 eV) and second excited state (Ar$^{**}$ - 13.1 eV) respectively. At 5 mTorr gas pressure, ionization rate from both ground state and excited states is nearly uniformly distributed over the plasma bulk. The maximum value (at 5 mTorr) of ionization rate from ground state is ~1.9×10$^{20}$ m$^{-3}$s$^{-1}$ (Figure 3 (a)), whereas the maximum ionization rate of the first excited state (Ar$^*$ - 11.6 eV) and second excited state (Ar$^{**}$ - 13.1 eV) are ~0.02×10$^{20}$ m$^{-3}$s$^{-1}$ (Figure 3 (b)) and ~0.07×10$^{20}$ m$^{-3}$s$^{-1}$ (Figure 3 (c)) respectively, which are substantially lower. The ionization rate profile is consistent with the centre peaked plasma density profile shown in Figure 2 (a). The spatially symmetric profile is responsible for equal flux on powered and grounded electrodes at 5 mTorr gas pressure. As the gas pressure increases, the ground state ionization rates in the bulk plasma decrease drastically and the peaks appear near to the powered and grounded electrodes. It is observed that the spatial asymmetry in the ground state ionization rate is highest at 75 mTorr gas pressure (figure 3 (a)) thus producing the highest flux asymmetry in the discharge, as mentioned earlier and shown in figure 2 (e). The peak ionization rate near to the grounded electrode is responsible for higher flux to it. The ionization rates from $Ar^*$ and $Ar^{**}$ continue to grow with increasing gas pressure with higher peak value near to the grounded electrode. It is noticed that the maximum ionization rate from $Ar^{**}$ is higher when compared to $Ar^*$. At 500 mTorr gas pressure the maximum ionization rate from $Ar^*$ and $Ar^{**}$ are ~$1.0\times10^{20}$ $m^{-3}s^{-1}$ and ~$1.28\times10^{20}$ $m^{-3}s^{-1}$ respectively, which is nearly the same as ground state peak ionization rate near to grounded electrode (figure 3 (a)) but spread over whole plasma volume thus producing higher plasma density due to multistep ionizations. This confirms that the enhanced plasma density and symmetric profile generated at high gas pressures are due to ionization from metastable states (Ar* and Ar**) in the plasma bulk. In other words, at higher gas pressures, the multistep ionization processes dominate compared to ground state or direct ionization.

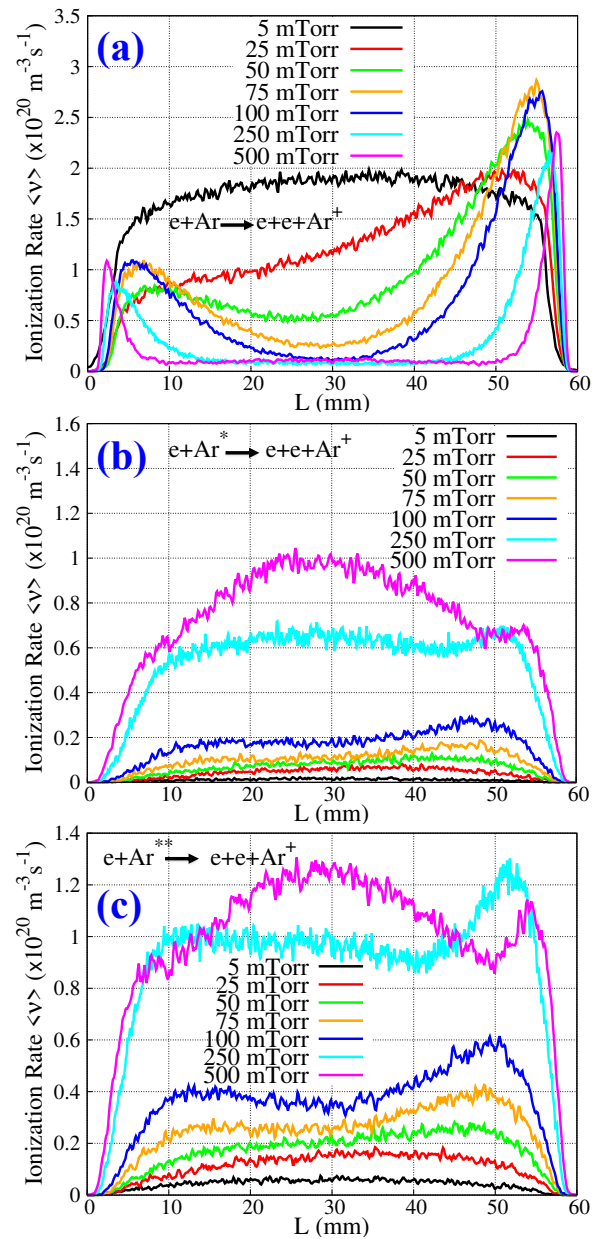

Figure 3: Time-averaged spatial profile of ionization rate from (a) ground state, (b) first excited state (Ar$^*$ - 11.6 eV) and (c) second excited state (Ar$^{**}$ - 13.1 eV) at different gas pressures from 5 mTorr to 500 mTorr. The current density and driving frequency are 50 A/m$^2$ and 27.12 MHz respectively.

A detailed investigation of the ionization asymmetry at different gas pressures is performed by analyzing the temporal behavior of ionizing collision rate within the RF-phase. Figure 4 shows the spatio-temporal evolution of ground state ionizing collision rate plotted for 2 RF periods at 3 different gas pressures(i.e.5 mTorr, 75 mTorr and 500 mTorr). The plotted values are averaged over 100 RF cycles after the simulation reached in the steady state condition. As shown in figure 4 (a), at 5 mTorr gas pressure, the ionization rate is distributed over the whole plasma volume and is slightly higher near the grounded electrode when compared to the powered electrode. This is consistent with the time-averaged ionization rate plotted in figure 3(a) where a slightly shifted profile towards grounded electrode is observed. The distributed ionization rate profile is attributed to the longer electron-neutral mean free path, whereas a slight increase in ionization rate near to the grounded electrode is due to sheath asymmetry (figure 2 (e)) and multiple ionization beams generated from the grounded sheath. Such types of electron beams are observed in low-pressure and at higher driving frequencies[9,68]. Due to low pressure and high frequency operation, multiple beams like structure in the ionizing collision rate are observed that are prominent near to the grounded sheath where the sheath width is higher. The injection of multiple ionization beam like structures are due to the multiple electron beam ejection from the expanding sheath edge[69-71]. As the gas pressure increases to 75 mTorr, the ionization profile becomes asymmetric and localized. More ionization is observed near the grounded electrode in comparison to the powered electrode. The observed localization is due to a decrease in electron-neutral collision mean free path that creates more ionization near to the expanding phase of sheath edges on both the electrodes. Furthermore, within RF-phase, the number of multiple ionization beams from the grounded sheath decreases. This is prominent at 500 mTorr gas pressure where the plasma is more localized near to the electrodes and multiple ionization beams from the grounded electrode merged and form a broad ionization region. This effect is attributed to smaller sheath widths. It is observed that the sheath width at the grounded electrode is reduced to ~ 2.1 mm at 500 mTorr gas pressure, which is ~60% lower than the sheath width at 5 mTorr gas pressure (~5.6 mm). A reduction in the sheath width is responsible for the slower sheath velocity and thus affects the generation of multiple ionization beams. Furthermore, it is also observed that the ionization in bulk plasma from the ground state is almost diminished at higher gas pressure.

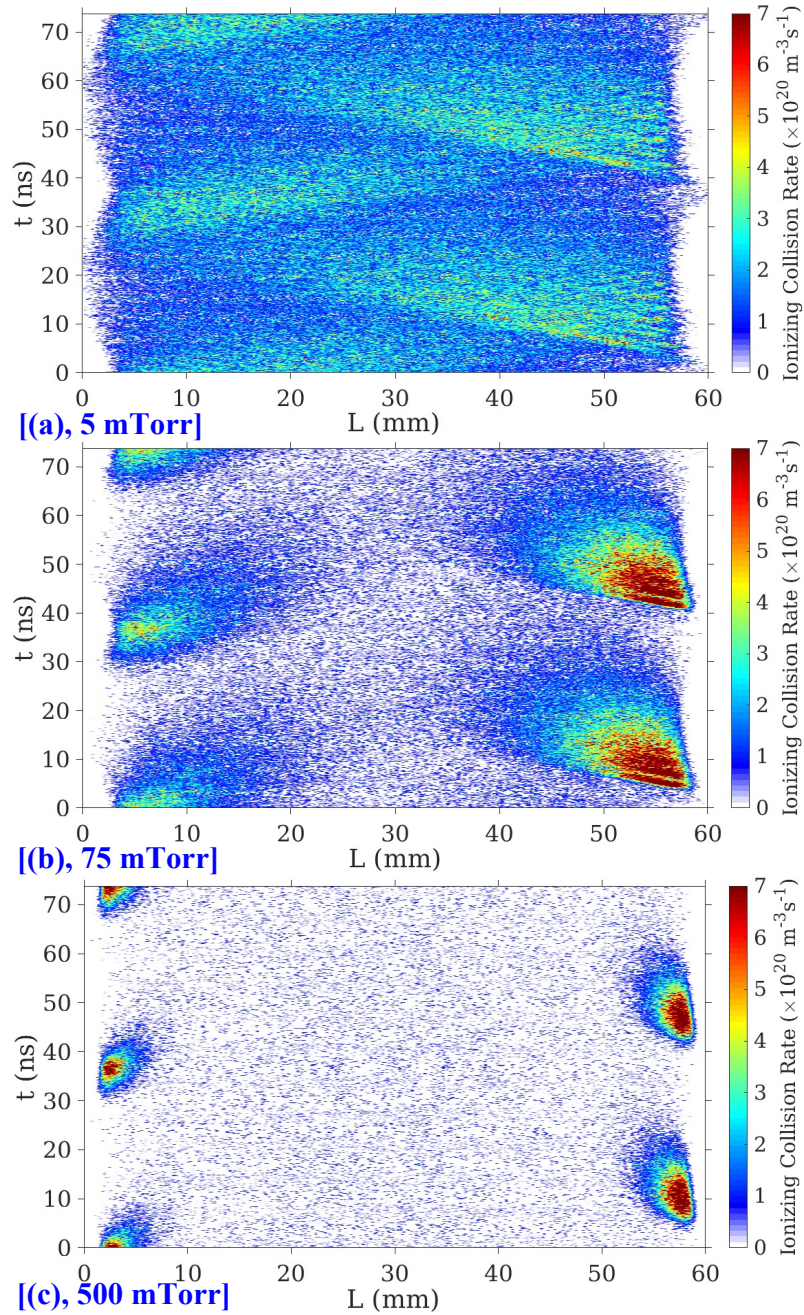

Figure 4: Spatio-temporal profile of ionizing collision rate plotted for 2 RF cycles at different gas pressures at (a) 5 mTorr, (b) 75 mTorr, and (c) 500 mTorr. The current density and driving frequency are 50 A/m$^2$ and 27.12 MHz respectively.

The suppression of higher harmonics, and thus a reduction in the generation of multiple ionization beams versus gas pressure, is studied by observing the electric field in the discharge. Figure 5 (a), 5 (b) and 5 (c) shows the spatio-temporal evolution of electric field at 5 mTorr, 75 mTorr and 500 mTorr gas pressures respectively for 2 RF periods. The plotted values are averaged over 100 RF cycles after the simulation reached in the steady state condition. As shown in figure 5 (a), the electric field is high in the sheath region and penetrating in the bulk plasma. The penetration of the electric field is caused by the generation of electron beams from the oscillating sheath edge position. It is evident from figure 5 (a) that the instantaneous sheath edge position is strongly modified with intense multiple higher harmonics. As the gas pressure increases to 75 mTorr, the sheath oscillations are suppressed and mainly present over the expanding phase of an RF sheath. At 500 mTorr gas pressure, the instantaneous sheath edge positions are nearly smooth, electric field is mostly confined in the sheath region and the electric field in the bulk plasma is mostly diminished. Figure 5 (d), 5 (e) and 5 (f) shows variation in electric field with time and its FFT (up to 2 GHz) at the centre of the discharge for 5 mTorr, 75 mTorr and 500 mTorr gas pressures respectively. At 5 mTorr gas pressure, the high frequency sheath oscillations correspond to the generation of distinct harmonic ($26^{th}$) along with the nearby harmonics that are observed in the bulk plasma. The corresponding electron plasma frequency ($f_{pe}$) at 5 mTorr gas pressure is ~635 MHz, therefore frequencies above $f_{pe}$ could penetrate in the discharge. In the present conditions, the observed harmonic frequency is ~705 MHz ($26^{th}$ harmonic). At low gas pressure, the maximum contribution of these harmonics reaches 40% (Figure 5(a)), suggesting power deposition through higher-order harmonics, whereas the fundamental is not present. As gas pressure increases to 75 mTorr, a significant drop, below 20%, in the higher harmonic contents is observed. In this case the dominant harmonic frequency is ~950 MHz ($35^{th}$ harmonics), whereas the corresponding $f_{pe}$ is ~940 MHz. For 500 mTorr gas pressure, no higher harmonics are observed in the electric field at the centre of the discharge. Furthermore, the average field intensity is nearly zero in this condition. The $f_{pe}$ is ~1750 MHz for electron density of $3.8 \times 10^{16}$ m$^{-3}$. The suppression of higher harmonics in the bulk plasma at higher pressures is attributed to smaller electron neutral collision mean free path (~0.03 cm at 500 mTorr) and plasma localization near to the sheath region as shown earlier.

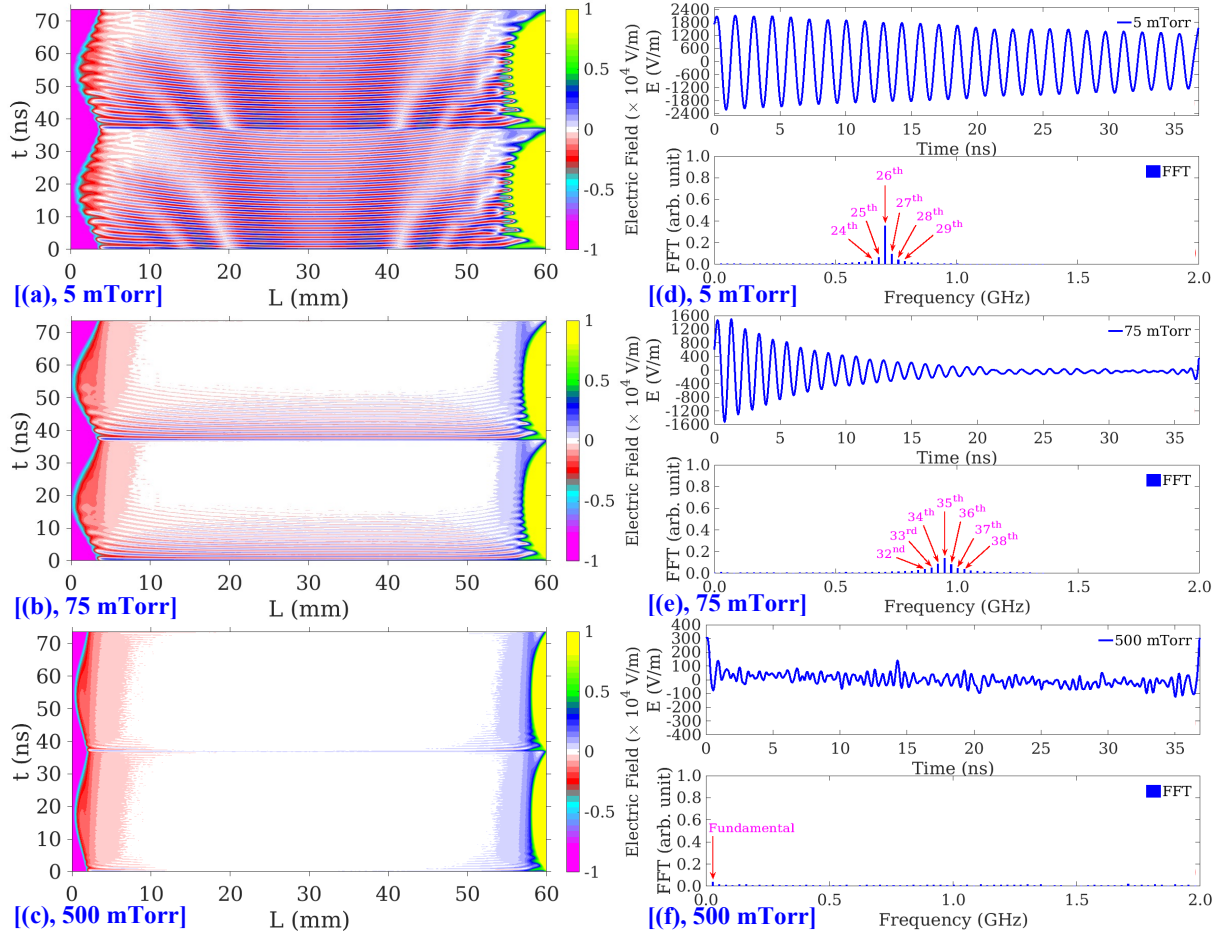

Figure 5: Spatio-temporal profile of electric field at (a) 5 mTorr, (b) 75 m Torr, (c) 500 mTorr and electric field and its FFT at the centre of the discharge for (d) 5 mTorr, (e) 75 mTorr, (f) 500 mTorr gas pressures. The current density and driving frequency are 50 A/m$^2$ and 27.12 MHz respectively.

A crucial parameter that controls the formation of metastable states, multistep ionization processes and plasma chemistry is the Electron Energy Distribution Function (EEDF). Figure 6 shows the EEDF at the centre of the discharge plotted for different gas pressures from 5 mTorr to 500 mTorr. As shown in figure 6, the EEDF is nearly bi-Maxwellian at 5 mTorr gas pressure with a large electron population ($5\times10^{15}$ m$^{-3}$) of low energy electrons (< 2 eV). A significant electron population (~$1\times10^8$ m$^{-3}$) is also present upto 90 eV energy. At low gas pressure, the bi-Maxwellian behavior of EEDF is well-known and caused by the heating of high-energy electrons through sheath, whereas the low-energy electrons are mostly confined in the bulk plasma due to ambipolar electric field[6,9,17]. Here, the significant population of high energy electrons is due to the penetration of multiple electron beams during the expanding sheath edge that are observed in both spatio-temporal ionizing collision rate (figure 4 (a)) and electric field penetration in the bulk plasma (figure 5 (a)). As the gas pressure increases, the population of low energy electrons increases, and the population of high energy electrons decreases. The EEDF changes its shape from bi-Maxwellian at 5 mTorr gas pressure to nearly Maxwellian at 25 mTorr gas pressure. As the gas pressure increases to 500 mTorr, a strong depletion of the high-energy electrons is observed decreasing below 25 eV. The depletion of high-energy electrons is due to increased collision rates that lead to a greater loss in electron energy near to the sheath region. Furthermore, at 500 mTorr gas pressure, heating of low energy electrons is observed that increases the electron population upto ~15 eV and the shape of the EEDF turn into convex nature with depleted tail. At high gas pressures, the heating of low-energy electrons is caused by the ohmic power deposition present in the plasma bulk[1,72]. An increase in the population of mid-energy range electrons generates higher metastable densities by electron impact excitation mechanism as shown in figure 2 (b) and 2 (c). Additionally, these electron populations also produce ionization from metastable states as presented in figure 3 (b) and 3 (c).

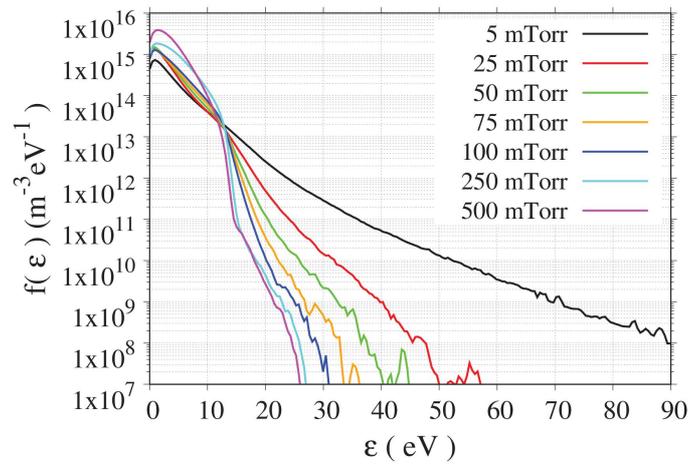

Figure 6: Electron Energy Distribution Function (EEDF) at the centre of the discharge for different gas pressures from 5 mTorr to 500 mTorr. The current density and driving frequency are 50 A/m$^2$ and 27.12 MHz respectively.

The above simulation studies are performed without considering the effect of secondary electron emissions, which is not significant in the present conditions due to low current density operation. We further investigate such effects by performing simulations including secondary electron emission (SEE). This study is carried out at 500 mTorr as the effect of SEE is dominant at high gas pressures, and by keeping the other operational parameters constant (27.12 MHz driving frequency and current density of 50 A/m$^2$). An ion induced secondary electron emission coefficient of 0.2 is considered. Figure 7 shows the plot of time-averaged (a) electron and ion densities profiles, (b) plasma potential profile, (c) flux distribution profile, to investigate and validate the spatial flux results as discussed earlier. As shown in figure 7 (a), the density profile with and without SEE remains at a centre peak and decreases towards the electrodes. An overall increase in the peak plasma density from 3.8×10$^{16}$ m$^{-3}$ to 4.5×10$^{16}$ m$^{-3}$ i.e., ~18% increase is noticed. However, by including the effect of SEE a change in the sheath dimensions on the electrodes are almost negligible. It is observed that the sheath width changes from 2.08 mm (without SEE) to 2.07 mm (with SEE) and 2.11 mm (without SEE) to 2.10 mm (with SEE) at powered and grounded electrode respectively. The plasma potential ($V_p$) profile shown in figure 7 (b) shows a decrease in the peak plasma potential from ~38 V to ~35 V by including SEE. On the other hand, the DC self-bias decreases by the proportionate value and thus the ion energy impinging on the power electrode remains unchanged. An increase in the bulk plasma density by including SEE is attributed to enhanced ionization rate, which is responsible for higher plasma density. A slight increment in the ionization rate is caused by the SEE that accelerates by the sheath electric field towards plasma bulk resulting in ionization near the sheath edges. Coming to the flux asymmetry, as shown in figure 7 (c), the flux profile is marginally affected by the SEE except for a small change in the flux density at the electrodes.

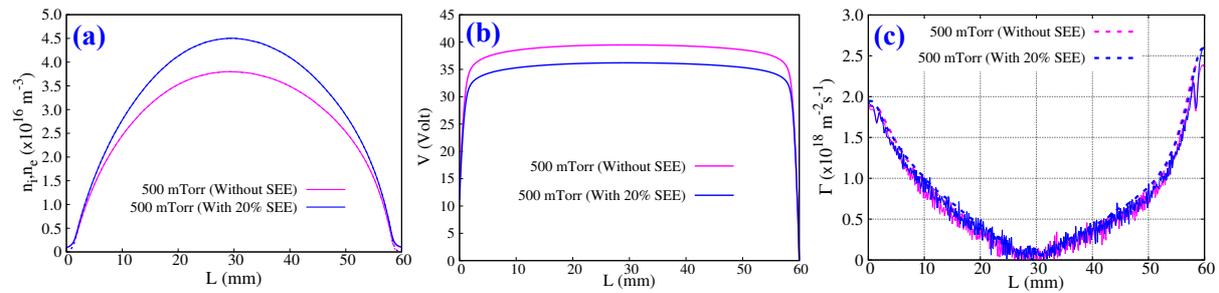

Figure 7: Time averaged (a) electron and ion density profile, (b) potential distribution, and (c) electron (line) and ion flux (dot) profile with and without SEE at 500 mTorr gas pressure. The current density and driving frequency are 50 A/m$^2$ and 27.12 MHz respectively.

By including SEE, the simulation results predict an ion flux of $\sim 1.95 \times 10^{18}$ m$^{-2}$s$^{-1}$ and $\sim 2.59 \times 10^{18}$ m$^{-2}$s$^{-1}$ at powered and grounded electrode respectively i.e., a ratio of $\sim 1.3$ ($\Gamma_{GE}/\Gamma_{PE}$) is observed, which is $\sim 1.27$ (figure 2 (e)) in case of without SEE. Thus, in the present operating conditions it is concluded that both the spatial flux and ion energy asymmetry are not significantly affected by including SEE.

Figure 8 (a)-(c) presents the time-averaged ionization collision rates from direct ionization, first excited state (Ar*), and second excited state (Ar**) with and without SEE at 500 mTorr, respectively. It is clear from Figure 8(a) that direct ionization is significantly higher near the sheath regions with a 20% SEE rate compared to without SEE. This occurs because the secondary electrons emitted from the electrodes are accelerated by the sheath electric field, leading to additional ionization near the sheath[72,73]. However, multistep ionization, specifically ionization from Ar* and Ar** metastable states, does not show significant differences between the presence and absence of SEE. We can conclude that the increase in density observed with 20% SEE is primarily due to the direct ionization process.

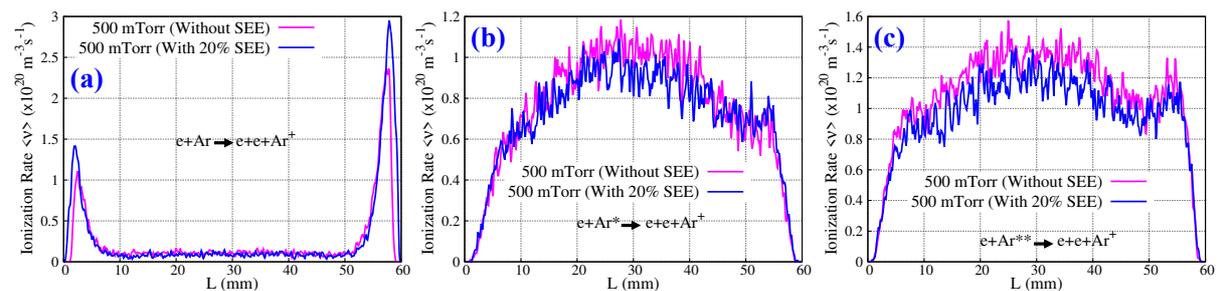

Figure 8: Time averaged (a) direct ionization (from ground state) collision rate, (b) ionization collision rate from first excited state (i.e. Ar$^*$), and (c) ionization collision rate from second excited state (i.e. Ar$^{**}$) with and without SEE at 500 mTorr gas pressure. The current density and driving frequency are 50 A/m$^2$ and 27.12 MHz respectively.

## 4. Summary and Conclusion

In a symmetric CCP discharge driven by a sawtooth-like waveform, the effects of gas pressure on plasma spatial asymmetry, metastable gas densities, ionization rates, higher harmonics generation, and electron energy distribution function are studied using the PIC simulation technique. At a low gas pressure (5 mTorr), the plasma distribution is substantially spatially asymmetric with a sheath ratio of ~1.35 (grounded to powered electrode, however, the flux asymmetry is marginal. As gas pressure increases, the plasma location near to the grounded electrode is observed. The flux asymmetry increases to ~ 30% at 75 mTorr gas pressure, whereas the sheath ratio drops to a minimum value of ~0.95. The DC self-bias in respect to plasma potential decreases from ~50 V at 5 mTorr to ~36 V at 75 mTorr. At higher pressure (500 mTorr), the profile, both flux and sheath ratios, turns into a nearly symmetric discharge. The metastable densities (Ar* and Ar**) follow the same trend as plasma density i.e., it is maximum in the plasma bulk at a low gas pressure and localized near to the sheath boundaries as gas pressure increases. The ionization rates from ground state and metastable states (multi-step ionization) indicate that the plasma density at a low gas pressure is mostly governed by the ground state ionization, whereas, at higher gas pressures the multi-step ionizations play an important role in the plasma density enhancement and symmetric spatial profile. A distinct higher harmonic is triggered in the plasma bulk at a low gas pressure due to high frequency oscillations on the instantaneous sheath edge position. At higher gas pressures, the instantaneous sheath edge behaves smoothly and the higher harmonics in the bulk plasma are suppressed. The electron energy distribution function is bi-Maxwellian at a low gas pressure, turning into nearly Maxwellian at intermediate pressure and then depleted high energy distribution is observed at high gas pressures. It is noticed that the effect of ion induced secondary electron emission on the plasma spatial asymmetry and flux asymmetry is marginal. From the simulation results, it is concluded that an optimum pressure condition is required for generating an appropriate plasma spatial and flux asymmetry.

**Acknowledgements**
This work was supported by the Science and Engineering Research Board (SERB), Core Research Grant No. CRG/2021/003536.

**Data Availability**

The data that support the findings of this study are available from the corresponding author upon reasonable request.